\documentclass[pra,twoside,twocolumn,showpacs]{revtex4-1}
\usepackage{amsmath}
\usepackage{amssymb}
\usepackage{amsfonts}
\newcommand{\vecg}{\boldsymbol}
\renewcommand{\vec}{\textbf}
\newcommand{\ket}[1]{|#1\rangle}

\newcommand{\bracket}[2]{\langle#1|#2\rangle}
\newcommand{\sg}[1]{\mathsf{#1}}

\begin{document}

\bibliographystyle{apsrev4-1}
\title{Spin operator in the Dirac theory}

\author{Pawe{\l}{} Caban}
\email{P.Caban@merlin.phys.uni.lodz.pl}
\author{Jakub Rembieli\'nski}
\email{jaremb@uni.lodz.pl}
\author{Marta W{\l}odarczyk}
\email{marta.wlodarczyk@gmail.com}

\affiliation{Department of Theoretical Physics, University of Lodz\\
Pomorska 149/153, 90-236 {\L}{\'o}d{\'z}, Poland}
\date{\today}

\begin{abstract}
We find all spin operators for a Dirac particle satisfying the
following very general conditions:
(i) spin does not
convert positive (negative) energy states into negative (positive)
energy states, (ii) spin is a pseudo-vector, and (iii) eigenvalues of the
projection of a spin operator on an arbitrary direction are independent
of this direction (isotropy condition).
We show that there are four such operators and all of them fulfill the
standard su(2) Lie algebra commutation relations. Nevertheless, only
one of them 
has a proper non-relativistic limit and acts in the same way on
negative and positive energy states. We show also that this operator
is equivalent to the Newton-Wigner spin operator and Foldy-Wouthuysen
mean-spin operator. We also discuss another operators proposed in the
literature. 
\end{abstract}
\pacs{03.65.Pm, 03.67.-a, 03.65.Ud}
\maketitle

\section{Introduction}

In recent years one can notice a renewal of interest in the
long-standing problem of the definition of a proper relativistic spin 
operator 
\cite{Terno2003,Ryder1999,PTW2013_spin_WKB,CRW_2013_Dirac_spin,%
BAKG2013}. 
One of the reasons of this renewal is the rapid development for
relativistic quantum information theory \cite{Czachor1997_1,PST2002,%
AM2002,CR2003_Wigner,ALMH2003,LD2003,CW2003,TU2003_1,PT2004_1,LD2004,%
LY2004,CR2005,KS2005,Czachor2005,CR2006,LMS2006,Caban2007_photons,%
BJ2008,HSZ2008,Caban2008,Moradi2008,%
CRW2008,DPV2009,LM2009,CRW2009,SHZ2009,%
Czachor2010,FBHH2010,CDKO_2010_fermion_helicity,HFGSH2011,%
CRWW_2011_hybrid,PVD2011,SV2011,SV2012,DV2012,SV2013}. 
In this context, especially important is the
simplest case, i.e., the spin operator for a Dirac particle. 
Many such operators have been proposed in the literature (see, e.g.,
\cite{Pryce1935,Pryce1948,FW1950,HW1963,Chakrabarti1963}). 
Nevertheless, it seems that the question which one of these is the best
is still open.

In this paper we address the problem. We formulate very general and
physically justified conditions which should be fulfilled by a
relativistic spin operator and classify all operators satisfying these
conditions. Our requirements are the following: (i) spin does not
convert positive (negative) energy states into negative (positive)
energy ones, (ii) spin is a pseudo-vector, and (iii) eigenvalues of the
projection of a spin operator in an arbitrary direction are independent
of this direction (isotropy). 
They are motivated by fundamental physical reasons: requirement
(i) follows from the fact that spin is an inner degree of freedom,
therefore it commutes with translations;
requirement (ii) is a consequence of the demand that spin should
transform in the 
same way as the total angular momentum; and requirement (iii) is
implied by the isotropy of space.
We show that there are four operators
fulfilling the above conditions. 

Note that we do not require any specific commutation relations for
components of a spin operator. However, it turns out that all four
operators satisfying our requirements fulfill the standard su(2) Lie
algebra commutation relations. 

Nonetheless, only one of those four operators has a proper
non-relativistic limit and satisfies the charge symmetry condition (acts
in the same way on positive and negative energy states). 
This operator turns out to be equivalent to the Newton-Wigner spin
operator and Foldy-Wouthuysen mean-spin operator. In our opinion it is
the best candidate for a relativistic spin operator for a Dirac
particle. 

We also compare operators we have found to various spin operators
presented in the literature. 

The paper is organized as follows.
In Sec.~\ref{sec:setup} we review briefly the abstract Dirac formalism
and its 
connection with the spin-1/2 unitary representation of the Poincar\'e
group. In Sec.~\ref{sec:rel_spin_operator} we discuss the relativistic
spin operator in 
the framework of the enveloping algebra of the Lie algebra of the
Poincar\'e group. 
We analyze in this context the influence of the overcompleteness of
the covariant basis on the identification of different forms of the
same operator. 
In Secs.~\ref{sec:spin_Bargmann} and
\ref{sec:spin_Dirac} we find all spin operators satisfying our
requirements in the Bargmann-Wigner and Dirac bases,
respectively. Section \ref{sec:review} is devoted to a comparison of
various spin operators presented in the literature.
Conclusions are given in Sec.~\ref{sec:conclusions}.

We use natural units with $\hbar=c=1$,
the Minkowski metric tensor $g^{\mu\nu}=\text{diag}(1,-1,-1,-1)$,
and adopt the convention $\varepsilon^{0123}=1$.

\section{The setup: abstract Dirac formalism}
\label{sec:setup}

A free spin-1/2 particle can be described in one of two
equivalent frameworks: a unitary representation of the Poincar\'e group
or with the help of the Dirac formalism. To establish the notation we
review here briefly basic 
facts concerning those two approaches. For the details we refer the
reader to, e.g., our previous paper \cite{CRW_2012_Dirac_formalism}.

\subsection{Bargmann-Wigner basis}

The space of states of a spin-1/2 particle,
$\mathcal{H}$, is the carrier space of the unitary representation of the
Poincar\'e group. To allow negative energies one takes as $\mathcal{H}$
the direct sum of two carrier spaces of unitary, irreducible
representations of the Poincar\'e group,
$\mathcal{H}=\mathcal{H}_+\oplus\mathcal{H}_-$, corresponding to
positive and negative energies, respectively. The space
$\mathcal{H}_{\sg{\epsilon}}$ ($\sg{\epsilon}=\pm1$, $+1$ corresponds
to positive energies; $-1$, to negative ones) is spanned by
eigenvectors of the four-momentum operators
 \begin{equation}
 \Hat{P}^\mu \ket{\sg{\epsilon}p,\sigma} = 
 \sg{\epsilon} p^\mu \ket{\sg{\epsilon}p,\sigma},
 \label{four-momentum_spin_basis}
 \end{equation}
where the spin index $\sigma=\pm1/2$.

We assume that vectors $\ket{\sg{\epsilon}p,\sigma}$ are normalized
covariantly
 \begin{equation}
 \bracket{\sg{\epsilon}^\prime
   p^\prime,\sigma^\prime}{\sg{\epsilon}p,\sigma} =
 2 p^0 \delta^3(\vec{p}^\prime-\vec{p}) 
 \delta_{\sg{\epsilon}^\prime \sg{\epsilon}} 
 \delta_{\sigma^\prime\sigma},
 \end{equation}
where
 \begin{equation}
 p^0=\sqrt{\vec{p}^2+m^2}.
 \end{equation}
Moreover, under Lorentz group action vectors
$\ket{\sg{\epsilon}p,\sigma}$ 
transform according to Wigner rotation
 \begin{equation}
 U(\Lambda) \ket{\sg{\epsilon}p,\sigma} = 
 \mathcal{D}(R(\Lambda,p))_{\lambda\sigma} 
 \ket{\sg{\epsilon}\Lambda p,\lambda},
 \label{transf_spin_basis}
 \end{equation}
where $\mathcal{D}$ stands for the unitary spin-1/2 representation of the
rotation group, $R(\Lambda,p)=L^{-1}_{\Lambda p} \Lambda L_p$ and
$L_p$ is a standard Lorentz transformation defined by: $L_p q = p$, 
$L_q=I$ with $q=(m,\vec{0})$. The basis defined by
Eq.~(\ref{four-momentum_spin_basis}) is called the Bargmann-Wigner or
spin basis.

\subsection{Dirac basis}

The main disadvantage of the Bargmann-Wigner basis is the
transformation law, (\ref{transf_spin_basis}), which implies that this
basis is not manifestly covariant.
Nevertheless, one can define
another basis of space $\mathcal{H}$,
 \begin{equation}
 \ket{\alpha, \sg{\epsilon}p} = 
 \sum_\sigma v^{\sg{\epsilon}}_{\alpha\sigma}(p) 
 \ket{\sg{\epsilon}p,\sigma}, 
 \label{cov_basis_def}
 \end{equation}
where we demand the following transformation rule: 
 \begin{equation}
 U(\Lambda) \ket{\alpha, \sg{\epsilon}p} = 
 S^{-1}(\Lambda)_{\alpha\beta}
 \ket{\beta, \sg{\epsilon}\Lambda p}.
 \label{transf_covariant_basis}
 \end{equation}
Here $S(\Lambda)$ designates the bispinor representation of
the Lorentz transformation $\Lambda$ and $\alpha$ is a bispinor
index. 

One can show that there exist coefficients
$v^{\sg{\epsilon}}_{\alpha\sigma}(p)$ such that
Eq.~(\ref{transf_covariant_basis}) holds; for details see,
e.g., \cite{CRW_2012_Dirac_formalism} and \cite{CRW_2013_Dirac_spin}.  
Intertwining matrices $v^{\sg{\epsilon}}_{\alpha\sigma}(p)$, related
to the Dirac amplitudes, fulfill relations
\begin{subequations}
 \begin{gather}
 v^{\sg{\epsilon}}(p) \bar{v}^{\sg{\epsilon}}(p) = 
 \sg{\epsilon} \Lambda_{\sg{\epsilon}}(p),\\
 v^{\sg{\epsilon}}(p) \bar{v}^{-\sg{\epsilon}}(p) = 
 - \sg{\epsilon} \Lambda_{\sg{\epsilon}}(p) \gamma^5,\\
 \bar{v}^{\sg{\epsilon}^\prime}(p) v^{\sg{\epsilon}}(p) =
 \sg{\epsilon} \delta_{\sg{\epsilon}^\prime\sg{\epsilon}} I_2,
  \end{gather}
\label{amplitudes_orthog_relations}%
\end{subequations}%
where $\bar{v}^{\sg{\epsilon}}(p)\equiv v^{\sg{\epsilon}\dag}(p)
\gamma^0$, $\gamma^\mu$ are Dirac matrices and the projectors
$\Lambda_{\sg{\epsilon}}(p)$ have the standard form
 \begin{equation}
 \Lambda_{\sg{\epsilon}}(p) = \frac{mI+\sg{\epsilon} p\gamma}{2m}. 
 \label{projectors_def}
 \end{equation}
We give the explicit form of the intertwining matrices
$v^{\sg{\epsilon}}_{\alpha\sigma}(p)$  and other useful formulas 
in Appendix~\ref{sec:app:Gamma}. 

Let us note that the spin basis $\{\ket{\sg{\epsilon}p,\sigma}\}$ is
complete, while the covariant basis is overcomplete; vectors
$\{\ket{\alpha,\sg{\epsilon}p}\}$ are constrained by the
Dirac condition.
Indeed, defining the Dirac operator
 \begin{equation}
 \Hat{D} \ket{\alpha,\sg{\epsilon}p} = 
 (\Hat{P}\gamma-mI)_{\alpha\beta} \ket{\beta,\sg{\epsilon}p} =
 (\sg{\epsilon}p\gamma-mI)_{\alpha\beta} \ket{\beta,\sg{\epsilon}p}, 
 \label{Dirac_eq_abs_1}
 \end{equation}
we get from
Eqs.~(\ref{cov_basis_def}) and (\ref{amplitudes_orthog_relations})
 \begin{equation}
 \Hat{D} \ket{\alpha,\sg{\epsilon}p} = 0;
 \label{Constraint_Dirac}
 \end{equation}
i.e.,
 \begin{equation}
 (p\gamma-\sg{\epsilon}mI)_{\alpha\beta} \ket{\beta,\sg{\epsilon}p} =
 0. 
 \label{Dirac_eq_abs}
 \end{equation}

This simple fact has very important consequences: the same abstract
operator can be represented by distinct matrices in Dirac theory
(cf.\ \cite{CRW_2013_Dirac_spin} and \cite{CR_2012_Comment_Spin_Dirac}).  
We discuss this point in detail below.

Now, let $\Hat{\Omega}$ be an $\Hat{P}^\mu$-dependent operator
acting in $\mathcal{H}$. In the Dirac
formalism $\Hat{\Omega}$ is represented by a $4\times4$ matrix with
matrix elements being functions of four-momentum operators
$\Hat{P}^\mu$: $\Hat{\Omega}=[\Omega(\Hat{P})_{\alpha\beta}]$. 
To determine the action of $\Hat{\Omega}$ on basis
vectors $\{\ket{\alpha,\sg{\epsilon}p}\}$ we have to take into account
that $\Hat{\Omega}$ in general can convert states with positive energy
into states with negative energy (and vice versa). Therefore, in the
abstract Dirac formalism we have
 \begin{equation}
 \Hat{\Omega} \ket{\alpha,\tilde{\sg{\epsilon}}p} = 
 \sum_{\sg{\epsilon}}
 \Omega_{\alpha\beta}^{\tilde{\sg{\epsilon}},\sg{\epsilon}}
 (\sg{\epsilon}p)  
 \ket{\beta,\sg{\epsilon} p},
 \label{end_Dirac_basis}
 \end{equation}
where
 \begin{equation}
 \Omega^{\tilde{\sg{\epsilon}}\sg{\epsilon}}
 (\sg{\epsilon} p) =
 \Lambda_{\tilde{\sg{\epsilon}}}(p) 
 \Omega(\sg{\epsilon} p)
 \Lambda_{\sg{\epsilon}}(p),
 \label{operator_projections}
 \end{equation}
with projectors $\Lambda_{\sg{\epsilon}}(p)$ defined in
Eq.~(\ref{projectors_def}). Note that in the above equation matrix
elements of $\Omega$ are functions of $\sg{\epsilon} p$, because
$\Hat{P}^\mu$ acting on state  
$\ket{\alpha,\sg{\epsilon} p}$ gives $\sg{\epsilon}
p^\mu \ket{\alpha,\sg{\epsilon} p}$.

Equation (\ref{end_Dirac_basis}) determines uniquely the action of an
operator $\Hat{\Omega}$ on the basis vectors. However, the matrix
$[\Omega(\Hat{P})_{\alpha\beta}]$
representing this operator is not determined uniquely. Indeed, let us
define the operator
 \begin{equation}
 \Hat{\Omega} + \Hat{A} \Hat{D},
 \label{representant_def}
 \end{equation}
where $\Hat{A}$ is an arbitrary $\Hat{P}^\mu$-dependent operator and
$\Hat{D}$ is given in Eq.~(\ref{Dirac_eq_abs_1}).
Then, by virtue of constraint (\ref{Constraint_Dirac}), we get
 \begin{multline}
 \Lambda_{\tilde{\sg{\epsilon}}}(p) 
 \Omega(\sg{\epsilon} p)
 \Lambda_{\sg{\epsilon}}(p) \\
 =
 \Lambda_{\tilde{\sg{\epsilon}}}(p) 
 \big[ \Omega(\sg{\epsilon} p) 
 + A(\sg{\epsilon}p) (\sg{\epsilon} p\gamma-mI) \big]
 \Lambda_{\sg{\epsilon}}(p),
 \end{multline}
meaning that operators 
$\Hat{\Omega}$ and $\Hat{\Omega} + \Hat{A} \Hat{D}$
represent the same abstract endomorphism.
In other words, any abstract endomorphism 
is represented by the whole class of operators
[Eq.~(\ref{representant_def})] acting in the same way on basis
vectors.  

Knowing the action of an operator $\Hat{\Omega}$ on basis
vectors $\ket{\alpha,\sg{\epsilon} p}$, i.e., having
Eq.~(\ref{end_Dirac_basis}), we can determine the matrix representing
$\Hat{\Omega}$. To this end, let us define the matrix
 \begin{equation}
 \tilde{\Omega}(p) = \sum_{\tilde{\sg{\epsilon}}\sg{\epsilon}} 
 \Omega^{\tilde{\sg{\epsilon}},\sg{\epsilon}}
 (\sg{\epsilon} p).
 \label{omega_tilde}
 \end{equation}
Of course, it holds that
 \begin{equation}
 \Lambda_{\tilde{\sg{\epsilon}}}(p) 
 \Omega(\sg{\epsilon} p)
 \Lambda_{\sg{\epsilon}}(p)
 =
 \Lambda_{\tilde{\sg{\epsilon}}}(p) 
 \tilde{\Omega}(\sg{\epsilon} p)
 \Lambda_{\sg{\epsilon}}(p).
 \end{equation}
Now, to obtain the matrix operator with matrix elements being
functions of four-momentum operators, we can use the equation
 \begin{equation}
 \frac{\Hat{P}^0}{|\Hat{P}^0|} \ket{\alpha,\sg{\epsilon}p} =
 \sg{\epsilon} \ket{\alpha,\sg{\epsilon}p},
 \label{def_operatora_epsilon}
 \end{equation} 
and write 
 \begin{equation}
 \tilde{\Omega}(p) = \tilde{\Omega}(\sg{\epsilon}(\sg{\epsilon}p))
 \quad \to \quad
 \tilde{\Omega}\Big( \frac{\Hat{P}^0}{|\Hat{P}^0|} \Hat{P} \Big).
 \label{omega_tilde_operator}
 \end{equation}
The operator $\Hat{\tilde{\Omega}}$ given by the matrix
$\tilde{\Omega}\Big( 
\frac{\Hat{P}^0}{|\Hat{P}^0|} \Hat{P} 
\Big)$ belongs to the class of (\ref{representant_def}). 

As a simple illustration of the above discussion let us consider the
energy operator
 \begin{equation}
 \Hat{H} = \Hat{P}^0.
 \end{equation}
For this operator we have
 \begin{equation}
 H(\sg{\epsilon}p) = \sg{\epsilon} p^0 I,
 \end{equation}
and, according to Eq.~(\ref{operator_projections})
 \begin{align}
 & H^{\sg{\epsilon}\sg{\epsilon}}(\sg{\epsilon}p) =
 \sg{\epsilon} p^0 \Lambda_{\sg{\epsilon}},\\
 & H^{-\sg{\epsilon},\sg{\epsilon}}(\sg{\epsilon}p) = 0.
 \end{align}
Now, Eq.~(\ref{omega_tilde}) implies that
 \begin{equation}
 \tilde{H}(p) = \frac{p^0}{m} p\gamma 
 = p^0 I + \frac{p^0}{m}(p\gamma-mI),
 \end{equation}
and from Eq.~(\ref{omega_tilde_operator}) we get
 \begin{equation}
 \Hat{\tilde{H}} 
 = \Hat{H} + \frac{\Hat{P}^0}{m} \Hat{D}.
 \end{equation}
It is also easy to see that the Dirac Hamiltonian is equivalent to 
$\Hat{H}$. The corresponding $\Hat{P}^\mu$-dependent matrix has the
form
 \begin{equation}
 H_D(\Hat{P}) = \gamma^0 (\Hat{\vec{P}}\cdot\vecg{\gamma}+mI)
 = \Hat{P}^0 I - \gamma^0 (\Hat{P}\gamma-mI).
 \end{equation}

\subsection{Interrelation between 
operators in the spin and Dirac bases}

Now, let $\Hat{\Omega}$ be an operator acting in space
$\mathcal{H}$. The action of $\Hat{\Omega}$ on the vector
$\ket{\sg{\epsilon}p,\lambda}$ in the Bargmann-Wigner basis can be
written in the form 
 \begin{equation}
 \Hat{\Omega} \ket{\tilde{\sg{\epsilon}}p,\lambda} = 
 \sum_{\sg{\epsilon}}
 \omega_{\sigma\lambda}^{\tilde{\sg{\epsilon}},\sg{\epsilon}}(p) 
 \ket{\sg{\epsilon} p,\sigma}.
 \label{end_spin_basis}
 \end{equation}
Thus, by virtue of Eqs.~(\ref{amplitudes_orthog_relations}) we obtain
in the Dirac basis
 \begin{equation}
 \Hat{\Omega} \ket{\alpha,\tilde{\sg{\epsilon}}p} = 
 \sum_{\sg{\epsilon}} 
 \big[ \sg{\epsilon} v^{\tilde{\sg{\epsilon}}}(p)
 {\omega^{\tilde{\sg{\epsilon}},\sg{\epsilon}}}^T(p)
 \bar{v}^{\sg{\epsilon}}(p) \big]_{\alpha\beta} 
 \ket{\beta,\tilde{\sg{\epsilon}}p}.
 \label{interrelation_general}
 \end{equation}
Therefore, by means of Eq.~(\ref{omega_tilde}),
the matrix $\tilde{\Omega}$ representing operator $\Hat{\Omega}$ in
the Dirac basis can be obtained from the following formula:
 \begin{equation}
 \tilde{\Omega}(p) = 
 \sum_{\tilde{\sg{\epsilon}}\sg{\epsilon}} 
 \big[ \sg{\epsilon} v^{\tilde{\sg{\epsilon}}}(p)
 {\omega^{\tilde{\sg{\epsilon}},\sg{\epsilon}}}^T(p)
 \bar{v}^{\sg{\epsilon}}(p) \big].
 \label{omega_tilde_Dirac_from_spin}
 \end{equation}

\subsection{Charge conjugation and parity}

One can also define the charge conjugation and parity
operators. 
On the level of quantum mechanics the charge conjugation operator
$\Hat{\mathbb{C}}$ is antiunitary \cite{IZ2006}.  
Thus, assuming that $\Hat{\mathbb{C}}$ commutes with Poincar\'e group
transformations, i.e.,
\begin{subequations}
 \begin{align}
 &[e^{ia_\mu \Hat{P}^\mu},\Hat{\mathbb{C}}]=0,
 \label{C_commutation_a}\\ 
 &[U(\Lambda),\Hat{\mathbb{C}}] = 0,
 \label{C_commutation_b}
 \end{align}%
\label{C_commutation}%
\end{subequations}
we get from Eq.~(\ref{C_commutation_a})
 \begin{equation}
 \Hat{\mathbb{C}} \Hat{P}^\mu =
 - \Hat{P}^\mu \Hat{\mathbb{C}}.
 \end{equation}
Therefore, $\Hat{\mathbb{C}}$
converts vectors with four-momentum $p$ into vectors with
four-momentum $-p$.
Thus, antiunitarity of $\Hat{\mathbb{C}}$ and
Eqs.~(\ref{transf_spin_basis}) and (\ref{C_commutation_b}) implies
 \begin{equation}
 \Hat{\mathbb{C}} \ket{\sg{\epsilon}p,\lambda} = 
 \sg{\epsilon} \xi_c (\sigma_2)_{\sigma\lambda}
 \ket{-\sg{\epsilon} p,\sigma},
 \label{charge_con_spin_def}
 \end{equation}
with $|\xi_c|=1$.

In the Dirac basis this operator acts as follows:
 \begin{equation}
 \Hat{\mathbb{C}} \ket{\alpha,\sg{\epsilon}p} =
 \xi_c \gamma^2_{\alpha\beta} 
 \ket{\beta,-\sg{\epsilon}p}.
 \end{equation}

We also use the parity operator $\mathbb{P}$ for which
$\mathbb{P}(p^0,\vec{p}) = (p^0,-\vec{p})\equiv p^\pi$. The action of
parity on vectors of the spin basis reads
\cite{CRW_2012_Dirac_formalism} 
 \begin{equation}
 \Hat{\mathbb{P}} \ket{\sg{\epsilon}p,\lambda} =
 \sg{\epsilon} \xi \ket{\sg{\epsilon}p^\pi,\lambda},
 \label{parity_spin_def}
 \end{equation}
where $|\xi|=1$.

The action of the parity operator on vectors of the Dirac basis reads
 \begin{equation}
 \Hat{\mathbb{P}} \ket{\alpha,\sg{\epsilon}p} = 
 \xi \gamma^0_{\alpha\beta} \ket{\beta,\sg{\epsilon}p^\pi}. 
 \label{parity_action_Dirac_basis}
 \end{equation}

\section{Relativistic spin operator}
\label{sec:rel_spin_operator}

There are different approaches to the definition of a relativistic
spin observable. First, one can try to split the total angular
momentum, $\Hat{\vec{J}}$, into the orbital part $\Hat{\vec{L}}$ and
spin part $\Hat{\vec{S}}$:
 \begin{equation}
 \Hat{\vec{J}} = \Hat{\vec{L}} + \Hat{\vec{S}}.
 \label{spin_splitting}
 \end{equation}
The total angular momentum is well defined \textit{via} generators of
the Lorentz group as
$\Hat{J}^i=\frac{1}{2}\varepsilon_{ijk}\Hat{J}^{jk}$. However, to find
$\Hat{\vec{L}}=\Hat{\vec{X}}\times\Hat{\vec{P}}$ one needs to know the
position operator $\Hat{\vec{X}}$. But a uniquely
defined relativistic position operator does not exist (in the literature);
different choices of 
$\Hat{\vec{X}}$ lead to different spin observables.

On the other hand, it is well known that in the unitary representation
of the Poincar\'e group there exists a well-defined spin-square
operator,
 \begin{equation}
 \Hat{\vec{S}}^2 = - \frac{1}{m^2} \Hat{W}^\mu \Hat{W}_\mu,
 \end{equation}
where $\Hat{W}^\mu$ is the Pauli-Lubanski four-vector
 \begin{equation}
 \Hat{W}^\mu = \frac{1}{2} \varepsilon^{\nu\alpha\beta\mu}
 \Hat{P}_\nu \Hat{J}_{\alpha\beta}
 \end{equation}
and $\Hat{J}_{\alpha\beta}$ are generators of the Lorentz group. Thus,
one can naturally try to define a spin operator as a linear function
of components of the Pauli-Lubanski four-vector. Of course,  for such a
function to constitute a spin operator it has to fulfill some
conditions which are believed to be the most important properties of
the spin observable. These conditions are the following:
(i) spin commutes with the four-momentum operators (this means that spin is
an inner degree of freedom)
 \begin{equation}
 [\Hat{S}^i,\Hat{P}^j]=0,
 \label{spin_cond_1}
 \end{equation}
(ii) spin components fulfill the standard su(2) Lie algebra
commutation relations 
 \begin{equation}
 [\Hat{S}^i,\Hat{S}^j] = i \varepsilon_{ijk} \Hat{S}^k,
 \label{spin_cond_2}
 \end{equation}
and (iii) spin transforms like a (pseudo)vector under rotations
 \begin{equation}
 [\Hat{J}^i,\Hat{S}^j] = i \varepsilon_{ijk} \Hat{S}^k.
 \label{spin_cond_3}
 \end{equation}
One can show \cite{CRW_2013_Dirac_spin} that the only operator
from the enveloping algebra of the Lie algebra of the Poincar\'e group
that is a linear function of the 
components of $\Hat{W}^\mu$ and has the properties
(\ref{spin_cond_1}) and (\ref{spin_cond_2}) has the following form:
 \begin{equation}
 \Hat{\vec{S}}_{NW} = \frac{1}{m} \left(
 \frac{|\Hat{P}^0|}{\Hat{P}^0} \Hat{\vec{W}}
 - \Hat{W}^0 \frac{\Hat{\vec{P}}}{|\Hat{P}^0|+m} 
 \right).
 \label{spin_NW_general}
 \end{equation} 
We use the notation $\Hat{\vec{S}}_{NW}$ because it appears that the
above operator can be obtained from Eq.~(\ref{spin_splitting}) if we
take the Newton-Wigner operator \cite{cab_NW1949} as a position operator
$\Hat{\vec{X}}$ (see,
e.g., \cite{CRW_2012_Dirac_formalism} and \cite{CRW_2013_Dirac_spin}). 
In the case where one considers only positive energies, the spin
operator, (\ref{spin_NW_general}), takes the form
 \begin{equation}
 \Hat{\vec{S}}_{NW} = \frac{1}{m} \left(
 \Hat{\vec{W}} - \Hat{W}^0 \frac{\Hat{\vec{P}}}{\Hat{P}^0+m} 
 \right).
 \label{spin_NW_positive}
 \end{equation}  
Let us stress that the spin operator, (\ref{spin_NW_positive}), was
discussed 
for the first time by Pryce in \cite{Pryce1935}. This spin operator
naturally arises also in quantum 
field theory (see, e.g., \cite{cab_BLT1969}). 

Operator (\ref{spin_NW_general}) transforms under Lorentz-group
action according to an operator Wigner rotation
\cite{CRW_2013_Dirac_spin}. 

The $\Hat{\vec{S}}_{NW}$ operator is defined in the enveloping algebra
of the Lie algebra of the Poincar\'e group. Therefore, it can be used
for a particle with arbitrary spin. In the following we discuss
this operator for a Dirac particle and compare it with various
operators proposed in the literature.

\section{Spin operator for a Dirac 
particle in the Bargmann-Wigner basis} 
\label{sec:spin_Bargmann}

Many different spin operators have been proposed for a Dirac particle.
We review them in Sec.~\ref{sec:review}, but first
we try to determine
the most general spin operator which fulfills very general physical
requirements. Namely, we only assume that spin is a pseudo-vector,
eigenvalues of the spin projection on an arbitrary direction
$\vec{a}$ are independent of $\vec{a}$ (isotropy condition),
and spin does not mix positive and negative energy states.
On the
level of quantum field theory this requirement is a consequence of the
superselection rule forbidding the superpositions of particle and
anti-particle states.

Because a matrix representation of
an abstract operator in the covariant (Dirac) basis is not unique
it is much more convenient to perform our analysis in the
Bargmann-Wigner basis.

A spin operator which does not mix positive and negative energy states
has the form:
 \begin{equation}
 \Hat{\vec{S}} \ket{\sg{\epsilon}p,\lambda} = 
 \vec{s}^{\sg{\epsilon}\sg{\epsilon}}(\vec{p})_{\sigma\lambda}
 \ket{\sg{\epsilon}p,\sigma},
 \label{spin_Bargmann_general}
 \end{equation}
[because $\vec{s}^{-\sg{\epsilon},\sg{\epsilon}}(\vec{p})=0$, compare
Eq.~(\ref{end_spin_basis})].  
We assume that $\Hat{\vec{S}}$ is a three-vector and does not change
under parity operation (i.e.\ $\Hat{\vec{S}}$ is a pseudo-vector). 
We have at our disposal only four independent three-vectors which can
be used for the construction of the matrix
$\vec{s}^{\sg{\epsilon}\sg{\epsilon}}(\vec{p})$: 
 \begin{equation}
 \boldsymbol{\sigma},\quad
 (\boldsymbol{\sigma}\cdot\vec{p})\vec{p},\quad 
 I\vec{p},\quad \boldsymbol{\sigma}\times\vec{p}. 
 \end{equation}
Now, the condition that $\Hat{\vec{S}}$ does not change under parity
 \begin{equation}
 \Hat{\mathbb{P}} \Hat{\vec{S}} \Hat{\mathbb{P}}^\dagger = \Hat{\vec{S}}
 \end{equation}
together with
Eqs.~(\ref{four-momentum_spin_basis}) and (\ref{parity_spin_def})
implies
 \begin{equation}
 \vec{s}^{\sg{\epsilon}\sg{\epsilon}}(\vec{p})
 = \vec{s}^{\sg{\epsilon}\sg{\epsilon}}(-\vec{p}). 
 \end{equation}
Thus, the most general form of the spin which is a
pseudo-vector and does not mix positive and negative energy states in
the Bargmann-Wigner basis is 
 \begin{equation}
 \vec{s}^{\sg{\epsilon}\sg{\epsilon}}(\vec{p})
 = \alpha(\vec{p},\sg{\epsilon}) \boldsymbol{\sigma} +
 \beta(\vec{p},\sg{\epsilon}) 
 (\vec{p}\cdot\boldsymbol{\sigma}) \vec{p},
 \label{spin_operator_general_spin_bas}
 \end{equation}
where $\alpha(\vec{p},\sg{\epsilon})$, $\beta(\vec{p},\sg{\epsilon})$
are scalar functions of $\vec{p}$. Now, eigenvalues of
$\vec{a}\cdot\Hat{\vec{S}}$ are 
independent of $\vec{a}$ iff
 \begin{equation}
 \beta(\vec{p},\sg{\epsilon}) = 0 \quad \text{or} \quad
 \vec{p}^2\beta(\vec{p},\sg{\epsilon}) +
 2\alpha(\vec{p},\sg{\epsilon})=0.  
 \end{equation}
Therefore, we arrive at two distinct possibilities:
 \begin{equation}
 \Hat{\vec{S}} \ket{\sg{\epsilon}p,\lambda} 
 = \alpha(\vec{p},\sg{\epsilon}) \boldsymbol{\sigma}^T_{\lambda\sigma} 
 \ket{\sg{\epsilon}p,\sigma},
 \label{spin_Bargman_general_1}
 \end{equation}
and
 \begin{equation} 
 \Hat{\vec{S}} \ket{\sg{\epsilon}p,\lambda} 
 = \alpha(\vec{p},\sg{\epsilon}) \Big[ \boldsymbol{\sigma}^T -
 \frac{2}{\vec{p}^2}
 (\vec{p}\cdot\boldsymbol{\sigma}^T) \vec{p}\Big]_{\lambda\sigma}
 \ket{\sg{\epsilon}p,\sigma}.
 \label{spin_Bargman_general_2}
 \end{equation}
Equations (\ref{spin_splitting}) and (\ref{C_commutation_b}) and the
antiunitarity of $\Hat{\mathbb{C}}$ give 
 \begin{equation}
 \Hat{\mathbb{C}} \Hat{\vec{S}} = 
 - \Hat{\vec{S}} \Hat{\mathbb{C}}.
 \label{C_S_commutation}
 \end{equation}
Thus, the charge symmetry implies that
spin should act in the same way on positive and negative energy
states.
Assuming that eigenvalues of
$\vec{a}\cdot\Hat{\vec{S}}$ are equal to $\pm1/2$, we have 
$\alpha(\vec{p},\sg{\epsilon})=\pm1/2$. 
Therefore, we finally get two operators
 \begin{align}
 \Hat{\vec{S}}_I \ket{\sg{\epsilon}p,\lambda} &
 = \frac{1}{2} \boldsymbol{\sigma}^T_{\lambda\sigma} 
 \ket{\sg{\epsilon}p,\sigma},
 \label{spin_Bargman_1}\\
 \Hat{\vec{S}}_{I\!I} \ket{\sg{\epsilon}p,\lambda} &
 = \frac{1}{2} \Big[ 
 \frac{2}{\vec{p}^2}
 (\vec{p}\cdot\boldsymbol{\sigma}^T) \vec{p}
 -\boldsymbol{\sigma}^T
 \Big]_{\lambda\sigma}
 \ket{\sg{\epsilon}p,\sigma},
 \label{spin_Bargman_2}
 \end{align} 
where we have chosen $\alpha(\vec{p},\sg{\epsilon})=1/2$ for the spin
operator given in Eq.~(\ref{spin_Bargman_general_1}) and
$\alpha(\vec{p},\sg{\epsilon})=-1/2$ for the operator given in
Eq.~(\ref{spin_Bargman_general_2}). Under these choices both
operators, (\ref{spin_Bargman_1}) and (\ref{spin_Bargman_2}), fulfill
the standard commutation relations (\ref{spin_cond_2}). This statement
is obvious for operator (\ref{spin_Bargman_1}); for operator
(\ref{spin_Bargman_2}) it is a consequence of the relation
 \begin{equation}
 \Big[ \frac{2}{\vec{p}^2}(\vec{p}\cdot\boldsymbol{\sigma})\vec{p} -
 \boldsymbol{\sigma} \Big]_i = R_{ij}(p) \sigma_j, 
 \end{equation}
where the matrix
 \begin{equation}
 R(p) = \frac{2 \vec{p}\otimes\vec{p}^T}{\vec{p}^2} - I_3
 \end{equation}
is a proper rotation.

It is worth stressing that we did not require any commutation relations
for the spin components. These relations have been received as a
by-product of more fundamental assumptions.

Both operators, $\Hat{\vec{S}}_I$ and $\Hat{\vec{S}}_{I\!I}$, fulfill
relation (\ref{C_S_commutation}).
However, $\Hat{\vec{S}}_I$ seems to be more advantageous because 
there is a problem with the limit
$\vec{p}\to0$ of the operator $\Hat{\vec{S}}_{I\!I}$ defined in
Eq.~(\ref{spin_Bargman_2}).
Namely, in this limit the first term on the right-hand side of
Eq.~(\ref{spin_Bargman_2}) does not vanish.
Nevertheless, we consider
$\Hat{\vec{S}}_{I\!I}$ for completeness.
For the same reasons we also discuss spin operators breaking the
charge symmetry, i.e.\ operators which act in a different
way on positive and negative energy states. Taking into account
Eqs.~(\ref{spin_Bargman_general_1}) and (\ref{spin_Bargman_general_2})
we get two such operators,
\begin{subequations}
\begin{align}
 & \Hat{\vec{S}}_{I\!I\!I} \ket{p,\lambda} 
 = \frac{1}{2} \boldsymbol{\sigma}^T_{\lambda\sigma} 
 \ket{p,\sigma},
 \label{spin_Bargman_3a}\\
 & \Hat{\vec{S}}_{I\!I\!I} \ket{-p,\lambda} 
 = \frac{1}{2} \Big[ 
 \frac{2}{\vec{p}^2}
 (\vec{p}\cdot\boldsymbol{\sigma}^T) \vec{p}
 -\boldsymbol{\sigma}^T
 \Big]_{\lambda\sigma}
 \ket{-p,\sigma}.
 \label{spin_Bargman_3b}
 \end{align}
\label{spin_Bargman_3}
\end{subequations}
and
\begin{subequations}
\begin{align}
 & \Hat{\vec{S}}_{I\!V} \ket{p,\lambda} 
 = \frac{1}{2} \Big[
 \frac{2}{\vec{p}^2}
 (\vec{p}\cdot\boldsymbol{\sigma}^T) \vec{p}
 - \boldsymbol{\sigma}^T
 \Big]_{\lambda\sigma}
 \ket{p,\sigma},
 \label{spin_Bargman_4a}\\
 & \Hat{\vec{S}}_{I\!V} \ket{-p,\lambda} 
 = \frac{1}{2} \boldsymbol{\sigma}^T_{\lambda\sigma} 
 \ket{-p,\sigma}.
 \label{spin_Bargman_4b}
 \end{align}
\label{spin_Bargman_4}
\end{subequations}
Obviously, these operators do not satisfy the charge symmetry
condition, Eq.~(\ref{C_S_commutation}).
We conclude that the operator $\Hat{\vec{S}}_I$ is the best one.

\section{Spin operator in the Dirac basis}
\label{sec:spin_Dirac}

Now, we find the form of the discussed spin operators in the Dirac
basis. 
If the action of the spin operator $\Hat{\vec{S}}$ in the spin basis
is given in Eq.~(\ref{spin_Bargmann_general}), then
Eq.~(\ref{interrelation_general}) implies, in the Dirac basis,
 \begin{equation}
 \Hat{\vec{S}} \ket{\alpha,\sg{\epsilon}p} =
 \vec{S}^{\sg{\epsilon}\sg{\epsilon}}(\sg{\epsilon}p)_{\alpha\beta}
 \ket{\beta,\sg{\epsilon}p},
 \end{equation}
with
 \begin{equation}
 \vec{S}^{\sg{\epsilon}\sg{\epsilon}}(\sg{\epsilon}p)
 = \sg{\epsilon} 
 v^{\sg{\epsilon}}(p) {\vec{s}^{\sg{\epsilon}\sg{\epsilon}}}^T(p) 
 \bar{v}^{\sg{\epsilon}}(p).
 \label{spin_matrix_formula}
 \end{equation}

By virtue of Eq.~(\ref{spin_matrix_formula}), we get, for the spin operators
defined in Eqs.~(\ref{spin_Bargman_1}) and (\ref{spin_Bargman_2}),
 \begin{multline}
 \vec{S}_{I}^{\sg{\epsilon}\sg{\epsilon}}(\sg{\epsilon}p) 
 = \frac{\gamma^5}{4m} \Big\{ 
 \sg{\epsilon} \Big[ 
 m \boldsymbol{\gamma} - \vec{p} \gamma^0
 + \frac{\vec{p}}{p^0+m}(\vec{p}\cdot\boldsymbol{\gamma})\Big]\\
 + \boldsymbol{\gamma} (p\gamma) - \vec{p}
 + \frac{\vec{p}}{p^0+m} \gamma^0 (\vec{p}\cdot\boldsymbol{\gamma})  
 \Big\},
\end{multline}
\begin{multline}
  \vec{S}_{I\!I}^{\sg{\epsilon}\sg{\epsilon}}(\sg{\epsilon}p)
  = \frac{\gamma^5}{4m} 
 \Big\{ \sg{\epsilon} \Big[
 \frac{\vec{p}}{p^0-m}(\vec{p}\cdot\boldsymbol{\gamma}) 
 - m \boldsymbol{\gamma} - \vec{p} \gamma^0
 \Big]\\
 + \vec{p} - \boldsymbol{\gamma} (p\gamma)
 - \frac{\vec{p}}{p^0-m} \gamma^0 (\vec{p}\cdot\boldsymbol{\gamma})  
  \Big\}.
\end{multline}

Furthermore, according to Eq.~(\ref{omega_tilde}),
we obtain the spin matrix
\textit{via} the formula
 \begin{equation}
 \vec{S}(p) = \sum_{\sg{\epsilon}} 
 \vec{S}^{\sg{\epsilon}\sg{\epsilon}}(\sg{\epsilon}p).
 \label{spin_matrix_def}
 \end{equation}

Thus, spin matrices representing operators $\Hat{\vec{S}}_{I}$ and
$\Hat{\vec{S}}_{I\!I}$ derived by means of the procedure
(\ref{omega_tilde_operator}) are
 \begin{equation}
 \vec{S}_I(\Hat{P})  = 
 \frac{\gamma^5}{2m} \Big\{  
 \frac{\Hat{P}^0}{|\Hat{P}^0|} \Big[
 \boldsymbol{\gamma} (\Hat{P}\gamma) - \Hat{\vec{P}} 
 \Big] +
 \frac{\Hat{\vec{P}} \gamma^0
   (\Hat{\vec{P}}\cdot\boldsymbol{\gamma})}{|\Hat{P}^0|+m}
 \Big\},
 \label{S_I_operator}
 \end{equation} 
and
 \begin{equation}
 \vec{S}_{I\!I}(\Hat{P})  = 
 - \frac{\gamma^5}{2m} \Big\{ 
 \frac{\Hat{P}^0}{|\Hat{P}^0|}\Big[
 \boldsymbol{\gamma} (\Hat{P}\gamma) - \Hat{\vec{P}}
 \Big] 
 + \frac{\Hat{\vec{P}} \gamma^0
   (\Hat{\vec{P}}\cdot\boldsymbol{\gamma})}{|\Hat{P}^0|-m}
 \Big\},
 \end{equation}
respectively. 

In the same way we find the matrix representation of 
the spin operators given
in Eqs.~(\ref{spin_Bargman_3}) and (\ref{spin_Bargman_4}). We get 
 \begin{equation}
 \vec{S}_{I\!I\!I}(\Hat{P}) = 
 \frac{\gamma^5}{2} \Big[  
 \boldsymbol{\gamma} - \frac{\Hat{\vec{P}}}{|\Hat{P}^0|^2-m^2}
 (\gamma^0+I) (\Hat{\vec{P}}\cdot\boldsymbol{\gamma})
 \Big],
 \end{equation}
and
 \begin{equation}
 \vec{S}_{I\!V}(\Hat{P}) = 
 \frac{\gamma^5}{2} \Big[  
 -\boldsymbol{\gamma} - \frac{\Hat{\vec{P}}}{|\Hat{P}^0|^2-m^2}
 (\gamma^0-I) (\Hat{\vec{P}}\cdot\boldsymbol{\gamma})
 \Big],
 \end{equation}
respectively.

Evidently, when we restrict ourselves to positive energy states,
operators $\Hat{\vec{S}}_{I}$ and $\Hat{\vec{S}}_{I\!I\!I}$
coincide.  The same statement holds for operators 
$\Hat{\vec{S}}_{I\!I}$ and $\Hat{\vec{S}}_{I\!V}$.

\section{Review of spin operators discussed 
in the literature}
\label{sec:review}

\begin{table*}
\begin{center}
\begin{ruledtabular}
\begin{tabular}{rll}
 & Traditional form & Form used in our paper \\
\hline
 1. &
 $\Hat{\vec{S}}_D = -\frac{1}{2} \Hat{\vecg{\Sigma}}$ &
 $\Hat{\vec{S}}_{D} = -\frac{1}{2} \gamma^5 \gamma^0 \vecg{\gamma}$ \\
 2. &
 $\Hat{\vec{S}}_{NW} = - \frac{|\Hat{P}^0|}{2m} \Hat{\vecg{\Sigma}}
 + \frac{\Hat{\vec{P}}
   (\Hat{\vec{P}}\cdot\Hat{\vecg{\Sigma}})}{2m(m+|\Hat{P}^0|)} 
 + \frac{i \Hat{P}^0}{2m|\Hat{P}^0|}
 \Hat{\vec{P}}\times\vecg{\alpha}$ & 
 $\Hat{\vec{S}}_{NW} = \frac{\gamma^5}{2m}
 \left\{ \frac{\Hat{P}^0}{|\Hat{P}^0|}
 \left[\vecg{\gamma}({\Hat{P}}{\gamma}) - \Hat{\vec{P}} \right]
 +\frac{\Hat{\vec{P}}}{m+|\Hat{P}^{0}|} \gamma^0
 (\Hat{\vec{P}} \cdot \vecg{\gamma})\right\}$\\
 3. & 
 $\Hat{\vec{S}}_{FW} = - \frac{1}{2} \Hat{\vecg{\Sigma}} -
 \frac{i\beta}{2|\Hat{P}^0|} \Hat{\vec{P}}\times \vecg{\alpha}+
 \frac{\Hat{\vec{P}} \times
 (\Hat{\vecg{\Sigma}} \times \Hat{\vec{P}})}{2|\Hat{P}^0|(m+|\Hat{P}^{0}|)}$&
 $\Hat{\vec{S}}_{FW} = - \frac{1}{2|\Hat{P}^0|} \gamma^5 \gamma^0 
 \left\{ m\vecg{\gamma}-(\Hat{\vec{P}}\cdot\vecg{\gamma}) \vecg{\gamma}+
 \frac{\Hat{\vec{P}}}{m+|\Hat{P}^{0}|)} (\Hat{\vec{P}}\cdot\vecg{\gamma})
 -\Hat{\vec{P}} \right\}$\\
 4. & 
 $\Hat{\vec{S}}_{C} = - \frac{m^2}{2\Hat{P}^{0^2}} \Hat{\vecg{\Sigma}} 
 -\frac{im\beta}{2\Hat{P}^{0^2}} \Hat{\vec{P}}\times \vecg{\alpha}-
 \frac{\Hat{\vec{P}}\cdot\Hat{\vecg{\Sigma}}}{2\Hat{P}^{0^2}} \Hat{\vec{P}}$ &
 $\Hat{\vec{S}}_{C} = - \frac{1}{2\Hat{P}^{0^2}} \gamma^5 \gamma^0
 \left\{m^2\vecg{\gamma}-m(\Hat{\vec{P}}\cdot\vecg{\gamma})
 \vecg{\gamma}-m\Hat{\vec{P}} 
 +\Hat{\vec{P}}(\Hat{\vec{P}}\cdot\vecg{\gamma})\right\}$\\
 5. &
 $\Hat{\vec{S}}_{F} = - \frac{1}{2} \Hat{\vecg{\Sigma}}
 -\frac{i\beta}{2m} \Hat{\vec{P}}\times \vecg{\alpha}$ & 
 $\Hat{\vec{S}}_{F} = \frac{1}{2} \gamma^5 \gamma^0
 \left\{- \vecg{\gamma} + \frac{1}{m} (\Hat{\vec{P}}\cdot\vecg{\gamma})
 \vecg{\gamma} + \frac{1}{m} \Hat{\vec{P}}\right\}$\\
 6. &
 $\Hat{\vec{S}}_{Ch} = 
 - \frac{1}{2} \Hat{\vecg{\Sigma}} 
 + \frac{i}{2m} \Hat{\vec{P}} \times \vecg{\alpha}
 - \frac{\Hat{\vec{P}} \times (\Hat{\vecg{\Sigma}} \times
 \Hat{\vec{P}})}{2m(m+|\Hat{P}^{0}|)}$&  
 $\Hat{\vec{S}}_{Ch} = \frac{1}{2m} \gamma^5
 \left\{ - \gamma^0 \left[ |\Hat{P}^0| \vecg{\gamma} 
 - \frac{\Hat{\vec{P}}}{m+|\Hat{P}^0|}
 (\Hat{\vec{P}} \cdot \vecg{\gamma}) \right]
 +(\Hat{\vec{P}} \cdot \vecg{\gamma}) \vecg{\gamma}
 +\Hat{\vec{P}} \right\}$\\
 7. &
 $\Hat{\vec{S}}_{P} = - \frac{1}{2} \beta \Hat{\vecg{\Sigma}}
 - \frac{i\alpha_3 \alpha_2 \alpha_1 (\beta+1) 
 \vecg{\alpha} \cdot \Hat{\vec{P}}}{2\Hat{\vec{P}}^2} \Hat{\vec{P}}$ &
 $\Hat{\vec{S}}_{P} = \frac{\gamma^5}{2}
 \left\{ \vecg{\gamma} - \frac{\Hat{\vec{P}}}{\Hat{P}^{0^2}-m^2}
 (\gamma^0+1) (\Hat{\vec{P}} \cdot \vecg{\gamma}) \right\}$
\end{tabular}
\end{ruledtabular}
\end{center}
\caption{Definitions of various relativistic spin operators which
  appear in the literature. In the
  left column we present the form of spin operators given in the
  literature. In the right column we present the equivalent form used in
  our calculations ($\vecg{\Sigma}=\frac{1}{2i}
  \vecg{\alpha}\times\vecg{\alpha}$, $\beta=\gamma^0$, 
  $\vecg{\alpha}=\gamma^0\vecg{\gamma}$). In this table the sign of
  operators is opposite to the standard choice. The reason is that we
  want to preserve the algebraic structure (commutation relations) on
  the level of abstract Hilbert space.
\label{table_1}}
\end{table*}

In this section we review and compare various relativistic spin
operators for a Dirac particle
discussed in the literature. We have collected their definitions 
in Table~\ref{table_1}. 
A few remarks are in order about the notation we used. 

The first operator presented in Table~\ref{table_1},
$\Hat{\vec{S}}_{D}$, is the standard Dirac spin operator.

The second operator in Table~\ref{table_1} is called the Newton-Wigner
operator and denoted $\Hat{\vec{S}}_{NW}$ because it can be
obtained from Eq.~(\ref{spin_splitting}) under the assumption that 
the position operator is the Newton-Wigner operator
\cite{cab_NW1949}.
However, the abstract form of the operator $\Hat{\vec{S}}_{NW}$
[Eq.~(\ref{spin_NW_positive})] was discussed for the first time by
Pryce in 1935 \cite{Pryce1935}. 
This spin operator has also been used in quantum 
information theory (see, e.g.,
\cite{Terno2003,CR2005,CR2006,CRW2009}). 

The third operator in Table~\ref{table_1} is called the Foldy-Wouthuysen
operator since it was obtained in the classical paper 
devoted to the Foldy-Wouthuysen transformations of the Dirac equation
\cite{FW1950}. In Ref.~\cite{FW1950} this operator is called the
``mean-spin operator''. 

The fourth operator in Table~\ref{table_1} was given by Pryce in
\cite{Pryce1948}. In recent years this operator has been used by
Czachor in the context of quantum information theory \cite{Czachor1997_1}.

The fifth operator, $\Hat{\vec{S}}_{F}$, corresponds to the classical
spin discussed by Frenkel \cite{Frenkel1926_1}. 
This operator was also considered
in \cite{Pryce1948} and \cite{HW1963}. 

The sixth operator was introduced by Chakrabarti in
\cite{Chakrabarti1963}. 

The seventh operator, denoted by us as $\Hat{\vec{S}}_{P}$, has been
discussed recently in \cite{BAKG2013}.  
Authors of \cite{BAKG2013} named it ``Pryce operator'' in spite of the
fact that we could not find it in the Pryce works
\cite{Pryce1935,Pryce1948}. 

As noted before, the matrix representing the abstract operator is
not defined uniquely in the Dirac basis. Therefore, to compare spin
operators presented in Table~\ref{table_1} we calculate their
projections on positive and negative energy subspaces
[Eq.~(\ref{operator_projections})] and their form in the spin basis.
The results are presented in Table~\ref{table_2}. 
This table allows us to analyze properties of the various spin
operators.

First, we see that the best candidate for the relativistic spin
operator for a Dirac particle is the Newton-Wigner operator,
$\Hat{\vec{S}}_{NW}$. This operator is equivalent to the
Foldy-Wouthuysen mean-spin, $\Hat{\vec{S}}_{FW}$. 
These operators do not convert positive (negative) energy states into
negative (positive) ones
($\vec{s}^{-\sg{\epsilon,\sg{\epsilon}}}_{NW}(p)=
\vec{s}^{-\sg{\epsilon,\sg{\epsilon}}}_{FW}(p)=0$) and fulfill the
fundamental isotropy condition. Moreover, both of these operators act
in the same way on positive and negative energy states 
[$\vec{s}^{\sg{\epsilon\sg{\epsilon}}}_{NW}(p) =
\vec{s}^{\sg{\epsilon\sg{\epsilon}}}_{FW}(p)$ are independent of
$\sg{\epsilon}$] and this action in the Bargmann-Wigner basis is given
by the standard Pauli matrices. In our classification 
$\Hat{\vec{S}}_{NW}$ and $\Hat{\vec{S}}_{FW}$ are equivalent to the
operator $\Hat{S}_I$ [Eqs.~(\ref{spin_Bargman_1}) and
(\ref{S_I_operator})]. 

All other operators presented in Table~\ref{table_1} have some
disadvantages. 

The standard Dirac spin operator, $\Hat{\vec{S}}_D$, is excluded
because it can convert
positive energy states into negative ones, and vice versa. 
Even its projection to positive energy states does not satisfy the
isotropy condition. 

The Frenkel operator, $\Hat{\vec{S}}_F$, 
is in fact the sum of (block-diagonal) projections of the Dirac spin
operator on 
positive and negative energy sectors. For this reason it does not
convert positive (negative) energy states into negative (positive)
energy states. However, it does not
satisfy the isotropy condition. 
The same flaw has the spin operator $\Hat{\vec{S}}_C$.

The Chakrabarti spin operator, $\Hat{\vec{S}}_{Ch}$, for positive
energy states reduces to the Newton-Wigner operator:  
$\vec{s}^{++}_{Ch}(p)=\vecg{\sigma}/2$, 
$\vec{s}^{-+}_{Ch}(p)=0$.
However, for
negative energy states its action is different. Moreover, it can
convert negative energy states into positive ones.

The last operator, $\Hat{\vec{S}}_P$ is simply equal to our spin
operator $\vec{S}_{I\!I\!I}$. Therefore, it does not satisfy the
charge symmetry condition. Moreover, its nonrelativistic limit is
ill defined.

\begin{table*}
\begin{center}
\begin{ruledtabular}
\begin{tabular}{rll}
 & Projections in the Dirac basis & Projections in the spin basis \\
\hline
 1. &
 $\vec{S}^{\sg{\epsilon}\sg{\epsilon}}_{D}(\sg{\epsilon}p)
 = \frac{\sg{\epsilon}}{2m} \gamma^5 
 (p^0 \vecg{\gamma} - \gamma^0 \vec{p}) \Lambda_{\sg{\epsilon}}$ &
 $\vec{s}^{\sg{\epsilon}\sg{\epsilon}}_{D}(p) = 
 \frac{p^0}{2m} \big[
 \vecg{\sigma} 
 - \frac{\vec{p} (\vec{p}\cdot\vecg{\sigma})}{p^0(m+p^0)}
 \big]$ \\
  &
 $\vec{S}^{-\sg{\epsilon},\sg{\epsilon}}_{D}(\sg{\epsilon}p)
 = \frac{\gamma^5}{2} \big\{ \frac{\sg{\epsilon}}{m} 
 (p^0 \vecg{\gamma} - \gamma^0 \vec{p}) 
 -\gamma^0 \vecg{\gamma} \big\} \Lambda_{\sg{\epsilon}}$ & 
 $\vec{s}^{-\sg{\epsilon},\sg{\epsilon}}_{D}(p) = 
 \frac{i}{2m} \vec{p}\times\vecg{\sigma}$,\\
\hline
 2. &
 $\vec{S}^{\sg{\epsilon}\sg{\epsilon}}_{NW}(\sg{\epsilon}p)
 =  \frac{\gamma^5}{2} 
 \big\{ \sg{\epsilon} \vecg{\gamma} 
 -\frac{\vec{p}}{m+p^{0}}
 \big[ \sg{\epsilon} \gamma^0 + 1 \big] \big\}
 \Lambda_{\sg{\epsilon}},$ & 
 $\vec{s}^{\sg{\epsilon}\sg{\epsilon}}_{NW}(p) 
 = \frac{1}{2} \vecg{\sigma}$,\\
 &
 $\Vec{S}^{-\sg{\epsilon},\sg{\epsilon}}_{NW} = 0$, & 
 $\vec{s}^{-\sg{\epsilon},\sg{\epsilon}}_{NW}(p) = 0$,\\
\hline 
 3. &
 $\vec{S}^{\sg{\epsilon}\sg{\epsilon}}_{FW}(\sg{\epsilon}p)
 =  \frac{\gamma^5}{2} 
 \big\{ \sg{\epsilon} \vecg{\gamma}
 - \frac{\vec{p}}{m+p^{0}}
 \big[ \sg{\epsilon} \gamma^0 + 1 \big] \big\}
 \Lambda_{\sg{\epsilon}},$ &
 $\vec{s}^{\sg{\epsilon}\sg{\epsilon}}_{FW}(p) = \frac{1}{2} \vecg{\sigma}$, \\
 &
 $\vec{S}^{-\sg{\epsilon},\sg{\epsilon}}_{FW} = 0$, &
 $\vec{s}^{-\sg{\epsilon},\sg{\epsilon}}_{FW}(p) = 0$, \\
\hline
 4. &
 $\vec{S}^{\sg{\epsilon}\sg{\epsilon}}_{C}(\sg{\epsilon}p)
 = \frac{\gamma^5}{2p^0}
 (\sg{\epsilon} m \vecg{\gamma} - \vec{p}) \Lambda_{\sg{\epsilon}},$ &
 $\vec{s}^{\sg{\epsilon}\sg{\epsilon}}_{C}(p) = 
 \frac{m}{2p^0} \big[
 \vecg{\sigma} 
 + \frac{\vec{p} (\vec{p}\cdot\vecg{\sigma})}{m(m+p^0)}
 \big]$, \\
 & 
 $\vec{S}^{-\sg{\epsilon},\sg{\epsilon}}_{C}(\sg{\epsilon}p)
 = 0 $ &
 $\vec{s}^{-\sg{\epsilon},\sg{\epsilon}}_{C}(p) = 0$, \\
\hline
 5. & 
 $\vec{S}^{\sg{\epsilon}\sg{\epsilon}}_{F}(\sg{\epsilon}p)
 = \frac{\sg{\epsilon}}{2m} \gamma^5  
 (p^0 \vecg{\gamma} -\gamma^0 \vec{p})
 \Lambda_{\sg{\epsilon}}$, &
 $\vec{s}^{\sg{\epsilon}\sg{\epsilon}}_{F}(p) = 
 \frac{p^0}{2m} \big[
 \vecg{\sigma} 
 - \frac{\vec{p} (\vec{p}\cdot\vecg{\sigma})}{p^0(m+p^0)}
 \big]$, \\
 & 
 $\vec{S}^{-\sg{\epsilon},\sg{\epsilon}}_{F}(\sg{\epsilon}p)
 = 0$, & 
 $\vec{s}^{-\sg{\epsilon},\sg{\epsilon}}_{F}(p) = 0$, \\
\hline
 6. &
 $\vec{S}^{\sg{\epsilon}\sg{\epsilon}}_{Ch}(\sg{\epsilon}p)
 = \frac{\gamma^5}{2} \big\{ \frac{(\sg{\epsilon}-1) p^0}{m}
 \big[ \frac{p^0}{m} \vecg{\gamma} 
 - \vec{p}(\frac{1}{m}\gamma^0 + \frac{1}{m+p^0}) \big]
 + \vecg{\gamma} - \frac{\sg{\epsilon}\vec{p}}{m+p^0} (1+\gamma^0)
 \big\} \Lambda_{\sg{\epsilon}}$, &
 $\vec{s}^{\sg{\epsilon}\sg{\epsilon}}_{Ch}(p) = 
 \frac{\sg{\epsilon}}{2}\vecg{\sigma} 
 + \frac{1-\sg{\epsilon}}{2m^2} 
 \big[ {p^0}^2  \vecg{\sigma} 
 - \vec{p} (\vec{p}\cdot\vecg{\sigma})\big]$, \\
 &
 $\vec{S}^{-\sg{\epsilon},\sg{\epsilon}}_{Ch}(\sg{\epsilon}p)
 = \frac{(\sg{\epsilon}-1) p^0}{2m^2} \gamma^5 \gamma^0 
 \big\{ \vec{p} + (m-p^0 \gamma^0) \vecg{\gamma} 
 \big\} \Lambda_{\sg{\epsilon}}$, &
 $\vec{s}^{-\sg{\epsilon},\sg{\epsilon}}_{Ch}(p) = 
 \frac{i(1-\sg{\epsilon})p^0}{2m^2} \vec{p}\times\vecg{\sigma}$, \\
\hline
 7. &
 $\vec{S}^{\sg{\epsilon}\sg{\epsilon}}_{P}(\sg{\epsilon}p)
 = \frac{\gamma^5}{2} 
 \big\{ \vecg{\gamma} - \frac{p^0-\sg{\epsilon}m}{p^{0^2}-m^2} 
 \vec{p} (\gamma^0+1) \big\} \Lambda_{\sg{\epsilon}} $, &
 $\vec{s}^{\sg{\epsilon}\sg{\epsilon}}_{P}(p)
 = \frac{1}{2} \big[ \sg{\epsilon} \vecg{\sigma} 
 + \frac{(1-\sg{\epsilon})}{\vec{p}^2} \vec{p} 
 (\vec{p}\cdot\vecg{\sigma}) \big]$, \\
 &
 $\vec{S}^{-\sg{\epsilon},\sg{\epsilon}}_{P}(\sg{\epsilon}p)
 = 0$, &
 $\vec{s}^{-\sg{\epsilon},\sg{\epsilon}}_{P}(p)
 = 0$.
\end{tabular}
\end{ruledtabular}
\end{center}
\caption{Summary of the properties of various relativistic spin
  operators defined in Tab.~\ref{table_1}.
\label{table_2}}
\end{table*}

\section{Conclusions}
\label{sec:conclusions}

We have formulated very general physical conditions which should be
fulfilled by a spin operator for a Dirac particle. These conditions
are the following: 
(i) spin converts positive energy states into positive energy states
and negative energy states into negative energy states;
(ii) spin is a pseudo-vector; and
(iii) eigenvalues of the projection of the spin operator on arbitrary
direction $\vec{a}$ are independent of $\vec{a}$ (isotropy condition).

We have found all spin operators fulfilling the above conditions;
there exist four such operators. We have also shown that components of
all four of these operators fulfill the standard su(2) Lie algebra
commutation 
relations. However, only one of those operators, $\Hat{S}_I$ has a
proper non-relativistic limit and fulfills the charge symmetry
condition. This operator turns out to be equivalent to the
Newton-Wigner spin operator and Foldy-Wouthuysen mean-spin operator. 

All other spin operators discussed in the literature do not fulfill at
least one of our fundamental assumptions. It is noteworthy that
when we restrict ourselves to the positive energy sector only, then
besides the
Newton-Wigner and Foldy-Wouthuysen mean-spin operators, also the
Chakrabarti operator fulfills our requirements.

In the literature operators depending on an experimental
device can also be met. For example, the operator used in 
\cite{PTW2013_spin_WKB} depends on the direction
of a magnetic field in the Stern-Gerlach apparatus. But
for ultrarelativistic particles one cannot use the Stern-Gerlach
apparatus to measure spin. Instead, Mott polarimetry or M{\o}ller
polarimetry is used.

\begin{acknowledgments}
This work was supported by the University of Lodz and by the Polish
Ministry of Science and Higher Education under Contract
No.~N~N202~103738. 
\end{acknowledgments}

\appendix
\section{Gamma matrices and amplitudes}
\label{sec:app:Gamma}

In explicit calculations we use the following representation
of Dirac gamma matrices:
 \begin{equation}
 \gamma^0=\left(\protect\begin{array}{cc}
 0 & I \\ I & 0
 \end{array}\right),\,\,
 {\boldsymbol{\gamma}}=\left(\protect\begin{array}{cc}
 0 & -{\boldsymbol{\sigma}} \\ {\boldsymbol{\sigma}} & 0
 \protect\end{array}\right),\,\,
 \gamma^5= 
 \left(\protect\begin{array}{cc} 
 I & 0 \\ 0 & -I
 \end{array}\right),
 \label{gamma_explicit}
 \end{equation} 
where $\boldsymbol{\sigma}=(\sigma_1,\sigma_2,\sigma_3)$ and
$\sigma_i$ are the standard Pauli matrices.
The explicit form of amplitudes $v^{\sg{\epsilon}}(p)$ reads
(cf.~\cite{CRW_2012_Dirac_formalism,CRW_2013_Dirac_spin})
 \begin{equation}
 v^{\sg{\epsilon}}(p) = 
 \frac{1}{2\sqrt{1+\frac{p^0}{m}}} 
 \left(\protect\begin{array}{c}
 I_2 + \frac{1}{m} p^\mu \sigma_\mu \\[1mm]
 \sg{\epsilon} (I_2 + \frac{1}{m} {p^\pi}^\mu \sigma_\mu)
 \protect\end{array}\right) \sigma_2,
 \label{amplitudes_explicit}
 \end{equation}
where $\sigma_0=I_2$. Amplitudes (\ref{amplitudes_explicit}) and gamma
matrices (\ref{gamma_explicit}) fulfill the following relations:
 \begin{gather}
 \gamma^5 v^{\sg{\epsilon}}(p) = v^{-\sg{\epsilon}}(p),
 \label{ampl_rel_1}\\
 \gamma^0 v^{\sg{\epsilon}}(p) = 
 \sg{\epsilon} v^{\sg{\epsilon}}(p^\pi),
 \label{ampl_rel_2}\\
 \gamma^2 {v^{\sg{\epsilon}}}^*(p) = 
 - \sg{\epsilon} v^{-\sg{\epsilon}}(p) \sigma_2,
 \label{ampl_rel_3}\\
 - \bar{v}^{\sg{\epsilon}}(p) = \bar{v}^{-\sg{\epsilon}}(p) \gamma^5.
 \end{gather}

The following useful formulas can be proved by direct calculation:
 \begin{gather}
 \bar{v}^{\sg{\epsilon}}(p) \gamma^\mu v^{\sg{\epsilon}}(p) =
 \frac{p^\mu}{m} I_2,
 \label{formula_1}
 \\
 \bar{v}^{\sg{\epsilon}}(p) \gamma^5 v^{\sg{\epsilon}}(p) = 0,
 \label{formula_2}\\
 \bar{v}^{\sg{\epsilon}}(p) \gamma^0 \boldsymbol{\gamma}
 v^{\sg{\epsilon}}(p) = - \frac{\sg{\epsilon}i}{m} 
 \vec{p}\times\vecg{\sigma}^T,
 \label{formula_2a}\\
 \bar{v}^{\sg{\epsilon}}(p) \gamma^0 (\vec{p}\cdot\boldsymbol{\gamma})
 v^{\sg{\epsilon}}(p) = 0,
 \label{formula_3}\\
 \bar{v}^{\sg{\epsilon}}(p) \vecg{\gamma} (\vec{p}\cdot\vecg{\gamma})
 v^{\sg{\epsilon}}(p) = -\sg{\epsilon}\vec{p} I_2 
 + \frac{\sg{\epsilon}ip^0}{m} 
 \vec{p}\times\vecg{\sigma}^T,
 \label{formula_2b}\\ 
 \bar{v}^{\sg{\epsilon}}(p) \gamma^5 \gamma^0 v^{\sg{\epsilon}}(p) =
 \frac{1}{m} (\vec{p}\cdot\boldsymbol{\sigma}^T),
 \label{formula_4}
  \end{gather}
 \begin{gather}
 \bar{v}^{\sg{\epsilon}}(p) \gamma^5 \gamma^0 \boldsymbol{\gamma}
 v^{\sg{\epsilon}}(p) =
 \frac{\sg{\epsilon}}{m} 
 \Big[ \frac{\vec{p} (\vec{p}\cdot\boldsymbol{\sigma}^T)}{m+p^0}
 - p^0 \boldsymbol{\sigma}^T \Big],
 \label{formula_5}\\
 \bar{v}^{\sg{\epsilon}}(p) \gamma^5 \gamma^0 
 (\vec{p}\cdot\boldsymbol{\gamma}) v^{\sg{\epsilon}}(p) =
 - \sg{\epsilon}
 (\vec{p}\cdot\boldsymbol{\sigma}^T),
 \label{formula_6}\\
 \bar{v}^{\sg{\epsilon}}(p) \gamma^5 \boldsymbol{\gamma} 
 v^{\sg{\epsilon}}(p) = 
 \frac{1}{m} \Big[ m \boldsymbol{\sigma}^T +
 \frac{\vec{p}(\vec{p}\cdot\boldsymbol{\sigma}^T)}{m+p^0} 
 \Big],
 \label{formula_7}\\
 \bar{v}^{\sg{\epsilon}}(p) \gamma^5 \boldsymbol{\gamma} 
 (\vec{p}\cdot\boldsymbol{\gamma})  v^{\sg{\epsilon}}(p) = 
 \frac{\sg{\epsilon}}{m} \Big[ \vec{p}^2 \boldsymbol{\sigma}^T -
 \vec{p}(\vec{p}\cdot\boldsymbol{\sigma}^T) \Big],
 \label{formula_8}\\
 \bar{v}^{\sg{\epsilon}}(p) \gamma^5 \gamma^0 \boldsymbol{\gamma} 
 (\vec{p}\cdot\boldsymbol{\gamma})  v^{\sg{\epsilon}}(p) = 
  - \frac{1}{m} \vec{p} (\vec{p}\cdot\boldsymbol{\sigma}^T).
 \label{formula_9}
 \end{gather}

The algebra of gamma matrices and Eq.~(\ref{projectors_def}) imply: 
\begin{gather}
 \Lambda_\sg{\epsilon} \gamma^\mu \Lambda_\sg{\epsilon} = 
 \frac{\sg{\epsilon}p^\mu}{m} \Lambda_\sg{\epsilon},\\
 \Lambda_{-\sg{\epsilon}} \gamma^\mu \Lambda_\sg{\epsilon} = 
 \Big(\gamma^\mu-\frac{\sg{\epsilon}p^\mu}{m}I\Big)
 \Lambda_\sg{\epsilon},\\
 \Lambda_{\sg{\epsilon}} \gamma^{5} \vecg{\gamma}
 \Lambda_{\sg{\epsilon}} = \gamma^5 
 \Big( \vecg{\gamma} - \frac{\sg{\epsilon}\vec{p}}{m}I \Big)
 \Lambda_{\sg{\epsilon}},\\
 \Lambda_{-\sg{\epsilon}} \gamma^{5} \vecg{\gamma}
 \Lambda_{\sg{\epsilon}} = \frac{\sg{\epsilon}\vec{p}}{m} \gamma^5 
 \Lambda_{\sg{\epsilon}},\\
 \Lambda_{\sg{\epsilon}} \gamma^{5} (\vec{p}\cdot\vecg{\gamma})
 \Lambda_{\sg{\epsilon}} = p^0 \gamma^5 
 \Big( \gamma^0 - \frac{\sg{\epsilon}p^0}{m}I \Big)
 \Lambda_{\sg{\epsilon}},\\
 \Lambda_{-\sg{\epsilon}} \gamma^{5} (\vec{p}\cdot\vecg{\gamma})
 \Lambda_{\sg{\epsilon}} = \frac{\sg{\epsilon}\vec{p}^2}{m} \gamma^5 
 \Lambda_{\sg{\epsilon}},
 \end{gather}
 \begin{gather}
 \Lambda_{\sg{\epsilon}} \gamma^{5} \gamma^0 \vecg{\gamma}
 \Lambda_{\sg{\epsilon}} =
 \frac{\sg{\epsilon}}{m} \gamma^5 (\gamma^0 \vec{p}-\vecg{\gamma}p^0)
 \Lambda_{\sg{\epsilon}},\\
 \Lambda_{-\sg{\epsilon}} \gamma^{5} \gamma^0 \vecg{\gamma}
 \Lambda_{\sg{\epsilon}} =
 \big( \gamma^{5} \gamma^0 \vecg{\gamma} - \frac{\sg{\epsilon}}{m} \gamma^5
 (\gamma^0 \vec{p}-\vecg{\gamma}p^0) \big) \Lambda_{\sg{\epsilon}},\\
 \Lambda_{\sg{\epsilon}} \gamma^{5} \gamma^0 (\vec{p}\cdot\vecg{\gamma})
 \Lambda_{\sg{\epsilon}}
 = \gamma^{5} \gamma^0 \vec{p}\cdot\vecg{\gamma}
 \Lambda_{\sg{\epsilon}},\\
 \Lambda_{-\sg{\epsilon}} \gamma^{5} \gamma^0 
 (\vec{p}\cdot\vecg{\gamma}) \Lambda_{\sg{\epsilon}} = 0,\\
 \Lambda_{\sg{\epsilon}} \gamma^5 \gamma^0
 \big((\vec{p} \cdot \vecg{\gamma}) \vecg{\gamma} + \vec{p}\big)
 \Lambda_{\sg{\epsilon}}=0,\\
 \Lambda_{-\sg{\epsilon}} \gamma^{5} \gamma^0
 \big((\vec{p}\cdot\vecg{\gamma}) \vecg{\gamma}+\vec{p}\big)
 \Lambda_{\sg{\epsilon}} = 
 \gamma^{5} \gamma^0
 \big((\vec{p}\cdot\vecg{\gamma}) \vecg{\gamma}+\vec{p}\big)
 \Lambda_{\sg{\epsilon}},\\
 \Lambda_{\sg{\epsilon}} \gamma^{5}
 \big((\vec{p}\cdot\vecg{\gamma}) \vecg{\gamma} + \vec{p}\big)
 \Lambda_{\sg{\epsilon}} =
 \frac{\sg{\epsilon}}{m} \gamma^5
 \big((\vec{p}\cdot\vecg{\gamma}) \vec{p}-\vec{p}^2\vecg{\gamma}\big)
 \Lambda_{\sg{\epsilon}},\\
 \Lambda_{-\sg{\epsilon}} \gamma^{5}
 \big((\vec{p}\cdot\vecg{\gamma}) \vecg{\gamma} + \vec{p}\big)
 \Lambda_{\sg{\epsilon}} =
 \frac{\sg{\epsilon}}{m} \gamma^5 \gamma^0 p^0
 \big(\vec{p}+(\vec{p}\cdot\vecg{\gamma}) \vecg{\gamma}\big)
 \Lambda_{\sg{\epsilon}}.
\end{gather}


%

\end{document}